\documentclass[prd, amsfonts, longbibliography, nofootinbib, amsmath, amssymb, twocolumn, superscriptaddress]{revtex4}
\usepackage{bm, graphics, graphicx, epsfig, soul, latexsym, hyperref, mathrsfs, multirow}
\usepackage{amsmath}

\usepackage[english]{babel}
\usepackage{natbib}
\usepackage{siunitx}
\usepackage{footmisc}
\usepackage{grffile}
\usepackage[usenames]{color}
\definecolor{darkgreen}{rgb}{0,0.5,0}

\usepackage{subfigure}
\usepackage{bigints}
\usepackage{comment}
\usepackage{float}
\usepackage{multirow}
\usepackage{placeins}

\hypersetup{
    bookmarks=true,         
    unicode=false,          
    pdftoolbar=true,        
    pdfmenubar=true,        
    pdffitwindow=false,     
    pdfstartview={FitH},    
    pdftitle={My title},    
    pdfauthor={Author},     
    pdfsubject={Subject},   
    pdfcreator={Creator},   
    pdfproducer={Producer}, 
    pdfkeywords={keyword1} {key2} {key3}, 
    pdfnewwindow=true,      
    colorlinks=true,       
    linkcolor=red,          
    citecolor=cyan,        
    filecolor=magenta,      
    urlcolor=darkgreen,           
    linktocpage=true
}
\usepackage{ulem}
\allowdisplaybreaks
\begin{document}

\title{Detectability of gravitational higher order modes in the third-generation era}
\author{Divyajyoti}
\email{divyajyoti@physics.iitm.ac.in}
\affiliation{Indian Institute of Technology Madras, Chennai, 600036, India}
\affiliation{Centre for Strings, Gravitation and Cosmology, Department of Physics, Indian Institute of Technology Madras, Chennai 600036, India}
\author{Preet Baxi}
\email{pbaxi1199@gmail.com}
\affiliation{Indian Institute of Technology Madras, Chennai, 600036, India}
\affiliation{Centre for Strings, Gravitation and Cosmology, Department of Physics, Indian Institute of Technology Madras, Chennai 600036, India}
\author{Chandra Kant Mishra}
\email{ckm@physics.iitm.ac.in}
\affiliation{Indian Institute of Technology Madras, Chennai, 600036, India}
\affiliation{Centre for Strings, Gravitation and Cosmology, Department of Physics, Indian Institute of Technology Madras, Chennai 600036, India}
\author{K. G. Arun}
\email{kgarun@cmi.ac.in}
\affiliation{Chennai Mathematical Institute, Siruseri, 603103, India}

%

\begin{abstract}
Detection of higher order modes of gravitational waves in third-generation (3G) ground-based detectors such as Cosmic Explorer and Einstein Telescope is explored. Using the astrophysical population of binary black holes based on events reported in the second gravitational wave catalog by Laser Interferometer Gravitational Wave Observatory (LIGO) and Virgo (GWTC-2), in conjunction with the  Madau-Dickinson model for redshift evolution of the binary black hole mergers, we assess the detectability of these higher order modes using a network consisting of three third-generation detectors. We find that the two subleading modes [(3,3) and (4,4)] can be detected in approximately 30\% of the population with a network signal-to-noise ratio of 3 or more, and for nearly 10\% of the sources, the five leading modes will be detectable. Besides, a study concerning the effect of binary's mass ratio and its orbital inclination with the observer's line-of-sight in detecting various modes is presented. For a few selected events of the LIGO-Virgo catalog, we identify the modes that would have been detected if a third-generation detector was operational when these events were recorded. We also compute the detectability of higher modes by Voyager and find that only $\sim 6$ and $2\%$ of the detectable population will have an associated detection of (3,3) and (4,4) modes, respectively. Observing these higher order modes in the 3G era would have a huge impact on the science possible with these detectors ranging from astrophysics and cosmology to testing strong-field gravity.

\end{abstract}

\date{\today}

\maketitle
\section{Introduction}
\label{sec:intro}
Since the discovery of the first gravitational wave (GW) event\,\citep{Abbott:2016blz} in 2015, Laser Interferometer Gravitational Wave Observatory (LIGO) Scientific Collaboration and Virgo collaboration have reported the detection of a total of 47 compact binary mergers with false alarm rate $<1\text{yr}^{-1}$\,\citep{LIGOScientific:2020ibl}. These include 44 binary black hole (BBH), two binary neutron star, and one possible neutron star-black hole (NS-BH) mergers. Additionally, independent analyses using the publicly available data\,\cite{LSC:GWTC-2-GWOSC} have confirmed these detections and have also added a few marginal BBH events to the LIGO-Virgo catalog\,\cite{Antelis:2018smo, Venumadhav:2019tad,Venumadhav:2019lyq,Nitz:2018imz,Nitz:2019hdf}. These observations have provided us with several new insights into astrophysics, cosmology, and fundamental physics (see for instance, Refs.\,\citep{TheLIGOScientific:2016htt, GBM:2017lvd, Abbott:2017xzu, LIGOScientific:2020tif, LIGOScientific:2020kqk}).

The two advanced LIGO (aLIGO) detectors \citep{TheLIGOScientific:2014jea} in the U.S. and advanced Virgo \citep{Acernese:2015gua} in Europe participated in the first two observing runs (named, O1 and O2), as well as for the first half of the third observation run (O3a). The Japanese detector KAGRA \citep{Akutsu:2020his} joined the LIGO-Virgo network briefly towards the end of the third observing run. The second part of the third observing run (O3b) was concluded in March 2020. 

While existing detectors are being upgraded towards LIGO A+ design \citep{LIGOScientific:2019vkc, McClelland:T1500290-v3}, beyond A+ upgrades (LIGO Voyager \citep{McClelland:T1500290-v3}) and the next-generation detector configurations \cite{Punturo:2010zz, Dwyer:2014fpa} have already been proposed and science studies are currently ongoing. Cosmic Explorer (CE) \citep{McClelland:T1500290-v3, Dwyer:2014fpa, Evans:2016mbw} and Einstein Telescope (ET) \cite{Punturo:2010zz, Hild:2010id} are two leading third-generation (3G) detector proposals. Both are expected to have a strain sensitivity that is an order of magnitude better than the second-generation (2G) detectors such as aLIGO and aVirgo, and a low-frequency cutoff in the range of {$1$-$5$\,Hz \cite{Chamberlain:2017fjl}}.

While BBH mergers are already the most frequently detected GW events \cite{LIGOScientific:2020ibl}, they are also among the strongest emitters of GWs and are possibly the cleanest\footnote{Near-monochromatic GW signals from isolated neutron stars too can be modeled very accurately in GR and with relative ease.} ones to model using analytical and numerical techniques in general relativity (GR) \cite{Th300}. If the gravitational waveforms are known precisely, one can use a well-known data-analysis technique called {\it matched filtering} \citep{Helstrom68} to extract signals from noisy detector output. 
The method involves cross-correlating the detector data with an accurate set of templates that closely mimic the form of expected signals and are computed in advance. The success of the method critically depends on how accurate these model predictions are. This requirement has driven the signal-modeling efforts over the past four decades by the gravitational wave community across the globe \cite{Blanchet:2013haa, Bishop:2016lgv, Sasaki:2003xr,Ajith:2007qp,Buonanno:2000ef, Poisson:2011nh, Foffa:2013qca}. 

The exact form of the signal depends on several intrinsic and extrinsic source parameters. Hence it is important that its theoretical predictions (templates) include all possible effects, neglect of which can potentially induce systematic biases in the measurement of source parameters or worse can even lead to nondetection of these signals. One such effect is the presence of nonquadrupole modes (also referred to as subdominant modes or higher order modes) in signals from compact binary systems which are asymmetric (unequal mass components) and/or whose orbital planes are not optimally inclined towards the Earth (face-off binaries). The effect of non-quadruple modes on the detection and parameter estimation for binary black holes have been studied extensively (see for instance Refs.~\cite{Tagoshi:2014xsa, Varma:2014jxa, Varma:2016dnf, Kalaghatgi:2019log, Payne:2019wmy, Lange:2018pyp}) and have now been included in a number of models that have been obtained by performing fits to numerical relativity simulations (see for instance, Refs.\,\cite{Mehta:2017jpq, Mehta:2019wxm, London:2017bcn, Khan:2018fmp, Khan:2019kot}), or following the effective-one-body approach (see Refs.\,\citep{Cotesta:2018fcv, Cotesta:2020qhw}). 

\subsection{Implications of higher order modes}
\label{subsec:hm-implications}

One of the most important consequences of including higher order modes into the gravitational waveforms can be linked to their sensitivity to frequencies that are inaccessible through the dominant (quadrupole) mode. Typically, including higher order modes into the waveforms will extend the GW spectrum to higher frequencies. For instance, inspiral for the dominant (quadrupole) mode ($\ell$=2, $m$=2 or simply the 22 mode) can be assumed to terminate at twice the orbital frequency at the last stable orbit ($f_{\rm LSO}$), while the same for a higher mode waveform including the $k$th harmonic will be visible until the GW frequency becomes $k\,f_{\rm LSO}$. The direct consequence of this is the increase in the mass reach of broadband detectors \cite{VanDenBroeck:2006qu,Arun:2007qv}. 

The higher order modes, through amplitude corrections to gravitational waveforms, also bring in new dependencies in terms of the mass ratio, component spins, and inclination angle into the gravitational waveforms; see for instance Ref.~\cite{VanDenBroeck:2006ar} (nonspinning case), and \cite{Arun:2008kb} (for spinning case). By including them into the waveforms, one is able to break the degeneracies present in the waveform, such as those between inclination angle and luminosity distance~\citep{Usman:2018imj}, and that between mass ratio and spins \citep{Ohme:2013nsa, Hannam:2013uu}. This proves to be a very useful tool when extracting source properties and finds numerous implications in astrophysics ~\cite{VanDenBroeck:2006ar, Arun:2007hu, Trias:2007fp, Babak:2008bu, Arun:2014ysa, Varma:2014jxa, OShaughnessy:2014shr, Graff:2015bba, Varma:2016dnf, Bustillo:2015qty, Bustillo:2016gid, Kumar:2018hml, Kalaghatgi:2019log, Chatziioannou:2019dsz, Shaik:2019dym, Purrer:2019jcp, LIGOScientific:2020stg, Abbott:2020khf}, cosmology~\cite{Arun:2007hu, Babak:2008bu, Borhanian:2020vyr} and fundamental physics\citep{Shaik:2019dym, Purrer:2019jcp}. For instance, inclusion of higher modes breaks the distance-inclination angle degeneracy, allowing for their improved measurements. While better measurements of the luminosity distance allow putting tighter bounds on cosmological parameters such as the Hubble constant \citep{Sathyaprakash:2009xt, Borhanian:2020vyr}, improved inclination angle estimates can lead to better modeling of off-axis gamma ray bursts\citep{Arun:2014ysa}. 

Further, the use of higher modes has been shown to improve the efficiency of parametrized tests of GR~\cite{Mishra:2010tp, Yunes:2009ke} and massive graviton tests~\cite{Arun:2009pq}. 
 A new test of GR based on the consistency of different modes of the gravitational waveform was proposed~\cite{Dhanpal:2018ufk,Islam:2019dmk, Shaik:2019dym} and performed on a few selected events from the O3a\,\cite{Capano:2020dix}. A multipolar null test of GR was also proposed in Refs.~\cite{Kastha:2018bcr,Kastha:2019brk} which would measure the contribution to the gravitational waveforms from various multipoles and test their consistency with the predictions of GR.
 Recently, it was shown that detection of higher modes can improve the early warning time and localization of compact binary mergers, especially NS-BH systems\citep{Kapadia:2020kss, Singh:2020lwx}. 

\subsection{Detection of higher modes by LIGO/Virgo}
\label{sec:det-hm}

It should be clear from the discussion above that higher order modes become relevant when the binary is not face-on and/or its components have very different masses. Additionally, the multipolar structure of the radiation field guarantees relatively weaker strengths of higher order modes compared to the dominant quadrupole mode. In other words, we are more likely to detect quadrupolar mode from near-equal mass/face-on binary compared to nonquadrupole modes from an unequal mass/face-off system. This detection bias makes it difficult to detect higher order modes in observed sources. It was only recently that LIGO/Virgo observations showed the presence of these modes in an unambiguous way. While there was a hint of higher mode presence in the data for the event GW170729\,\citep{Chatziioannou:2019dsz}, clear evidence of a higher order mode was found during the analysis of two events namely, GW190412~\citep{LIGOScientific:2020stg} and GW190814~\citep{Abbott:2020khf}, both highly asymmetric in component masses. Further, for about six events (all observed during the first part of the third observing run of the LIGO-Virgo network), the inclusion of higher modes in waveform models was found to improve the parameter estimation accuracies~\citep{Abbott:2020tfl, LIGOScientific:2020ibl}, hinting at their presence.

\subsection{Motivation for the present work}
\label{sec:motiv}

As ground-based detectors improve their sensitivities over the next few years, they are going to detect more massive and more distant BBHs, should they exist. The increased mass reach is mostly due to the improved lower cutoff frequency of these detectors which may be as low as a few Hz.\footnote{Note that inclusion of higher modes also improves detector's mass reach as discussed above in Sec.~\ref{subsec:hm-implications}.}

The increased distance reach is due to the improved sensitivity at different frequency bands. Going by the present estimates,  these observations would definitely unravel more asymmetric binary systems many of which may not be face-on. This should facilitate detections of several of the subdominant modes by the next-generation detectors. As these higher modes would very likely bring in improvements to the parameter estimation in various contexts, a study of their detectability is a very important first step towards understanding the impact they will have on GW science. This forms the context of the present work where we quantify the detectability of nonquadrupolar modes using a network of future ground-based gravitational wave detectors.

Our study on a 3G detector network, using quasicircular, nonprecessing higher mode waveforms of Ref.~\cite{London:2017bcn}, finds that about 33\% of the population will detect the subleading mode, $\ell$=3, $m$=3 (or simply the 33 mode) and $\sim$28\% of the population will detect the $\ell$=4, $m$=4 (or 44 mode) mode, in addition to the dominant 22 mode. Further, for about 10\% of the population, it is possible to detect five leading spherical harmonic modes ({\it i.e.} 22, 21, 33, 32, and 44). These should have a profound impact on the planned astrophysics, cosmology, and fundamental physics using these detectors. 

The layout of the paper is as follows. Section\,\ref{sec:method} includes details of the waveform employed and our choice of detector network(s) used in the analysis. We start Sec.~\ref{sec:fiducial} by discussing the detection criteria (used throughout the paper) followed by results of a study concerning the detectability of higher order modes in the mass ratio ($q$) and inclination angle ($\iota$) plane. Additionally, detection of higher modes in selected GWTC-2 events, assuming a 3G detector was operational during the O3a run of LIGO and Virgo, is explored. Section\,\ref{sec:pop} presents the results of a full population study (based on the observed BBH population reported in \cite{LIGOScientific:2020ibl}) using a 3G detector network, along with a comparison study with a network of 2G detectors and their future upgrades.

\section{Waveforms and detector networks}
\label{sec:method}

\subsection{Spin-weighted spherical harmonic basis and the higher mode waveform structure}
\label{sec:wf}

Multipolar decomposition of the gravitational waveform is a convenient tool to represent the gravitational radiation from systems like compact binary mergers~\cite{Thorne:1980ru} and helps immensely in handling the nonlinearities of GR in the perturbative approaches to GR such as PN theory (see Ref.~\citep{Blanchet:2013haa} for a detailed review). Symmetric trace-free tensors and spin-weighted spherical harmonics provide two equivalent bases for such a decomposition (see for instance Refs.~\citep{Kidder:2007rt, Blanchet:2008je}). The latter has been more popular recently due to the extensive use of it by the numerical relativity community, as it provides a natural basis for extracting the waveform from numerical simulations (see for instance Ref.~\citep{Mroue:2013xna}).

The GW strain can be expressed as a linear combination of different modes defined using a basis of spin-weighted spherical harmonics of weight $-2$ as follows \citep{Goldberg:1966uu} 
\begin{equation}
    h(t,\overrightarrow\lambda,\Theta,\Phi) = \sum_{\ell \geq 2}\sum_{-\ell \leq m \leq \ell} h^{\ell m}(t,\overrightarrow\lambda) Y^{\ell m}_{-2}(\Theta, \Phi).
\label{eq:sph}
\end{equation}

\noindent Here, $t$ denotes the time coordinate, the intrinsic parameters like masses and spins are denoted by $\overrightarrow\lambda$, and ($\Theta$, $\Phi$) are the spherical angles in a source-centered coordinate system with total angular momentum along the $z$ axis. A number of waveforms, both numerical and phenomenological, have been developed which include higher modes\citep{Mehta:2017jpq, London:2017bcn, Blackman:2017pcm, Khan:2018fmp, Khan:2019kot, Mehta:2019wxm, Cotesta:2018fcv, Varma:2018mmi, Varma:2019csw, Rifat:2019ltp, Nagar:2020pcj, Garcia-Quiros:2020qpx, Cotesta:2020qhw, Ossokine:2020kjp, Pratten:2020ceb, Nagar:2020xsk, Foucart:2020xkt, Nagar:2021gss, Liu:2021pkr}. Many of these waveforms are incorporated in the LSC Algorithm Library Suite (\textsc{LALSuite})\citep{lalsuite}.
These waveforms make use of the analytical and semianalytical treatment of the compact binary dynamics within the PN~\cite{Blanchet:1995ez, Blanchet:1995fg, B96, BIJ02, BFIJ02, BDEI04, BIWW96, Arun:2004ff, Blanchet:2008je, Kidder:1995zr, BBuF06, Porto:2008jj, Porto:2008tb, Arun:2009pq, Porto:2010zg, MBFB2012, BMFB2012, BFH2012, M3B2013, Mishra:2016whh} and effective-one-body~\cite{BuonD98, Buonanno:2000ef, Damour:2000we, Damour:2001tu, Damour:2015isa, Goldberger:2004jt, Sennett:2019bpc} frameworks as well as of numerical relativity (NR) simulations (see Ref.~\citep{Boyle:2019kee} for a recent update on NR waveform catalog by the SXS collaboration and Ref.~\citep{Jani:2016wkt} by the Georgia Tech group. Both catalogs are publicly available; see also \cite{Brugmann:2008zz}). A comparison between different numerical relativity schemes leading to simulations of BBH spacetimes can be found in Refs.~\cite{Ajith:2012az,Hinder:2013oqa}.

 For our study, we choose to work with an inspiral-merger-ringdown waveform model of Phenom family for BBHs in quasicircular orbits including the effect of higher order modes and non-precessing spins (coded up in \textsc{LALSuite} with the name \texttt{IMRPhenomHM}) \citep{London:2017bcn}.  In addition to the dominant 22 mode this model can be used to extract other subdominant gravitational wave modes (21, 33, 32, 44, and 43) and is calibrated for the mass ratios ($q=m_1/m_2$; $m_1>m_2$) up to $18$, and component  dimensionless spin magnitudes up to $0.85$ (up to 0.98 for equal mass case) (see \citep{London:2017bcn} for details).
 
\subsection{Detector networks}
\label{sec:det-network}

For our study, we consider network(s) consisting of two kinds of 3G detectors: CE and ET. CE will be similar in layout to the current LIGO detectors, with two arms at a right angle to each other, forming an L shape. The length of these two arms is proposed to be 40\,km each, which is 10 times longer than the advanced LIGO detector. ET, on the other hand, will have a different layout. It will consist of three arms forming an equilateral triangle. Each arm will have a length of 10\,km, and the whole setup is underground. Both detectors are expected to achieve a sensitivity that is roughly an order of magnitude better than the current 2G detectors (aLIGO), on average, and a low-frequency sensitivity in the range 1-5\,Hz \cite{Chamberlain:2017fjl}. 

\begin{figure}[t]%
\centering
\includegraphics[trim=10 10 10 10, clip, width=\linewidth]{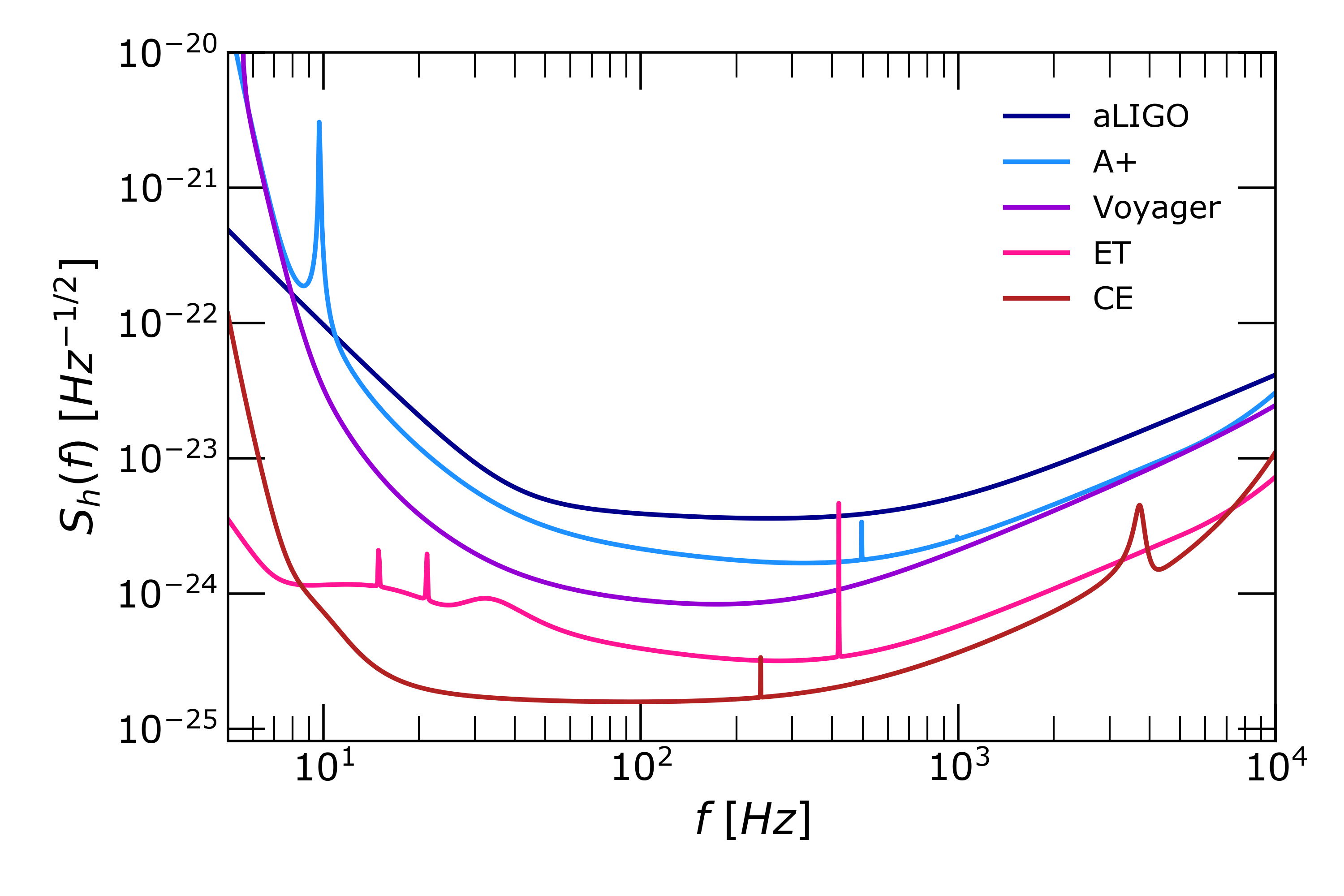}
\caption{Detector sensitivity curves for various detectors considered in this analysis. In addition, sensitivity for aLIGO is also shown for comparison.}
\label{fig:PSDs}
\end{figure}

Further, we also compare the detection of higher modes in the 3G network with that in the upgraded 2G networks with LIGO A+ configuration\,\citep{LIGOScientific:2019vkc, McClelland:T1500290-v3} and LIGO Voyager \citep{McClelland:T1500290-v3}. Both LIGO A+ and LIGO Voyager are expected to have an overall improved sensitivity compared to that of current generation detectors (see Fig.\,\ref{fig:PSDs}).    

\begin{table*}[]
\begin{tabular}{|c|c|c|c|c|c|}
\hline
Label & Location                   & Latitude & Longitude & Orientation & Type(s)         \\ \hline
L     & Louisiana, USA             & 0.53     & -1.58     & -1.26       & CE/A+/Voyager \\ \hline
H     & Washington, USA            & 0.81     & -2.08     & -2.51       & CE/A+/Voyager \\ \hline
V     & Cascina, Italy             & 0.76     & 0.18      & 2.8         & CE/A+/Voyager \\ \hline
A     & New South Wales, Australia & -0.59    & 2.53      & 0.78        & CE              \\ \hline
E     & Cascina, Italy             & 0.76     & 0.18      & 2.8         & ET              \\ \hline
\end{tabular}
\caption{The detector locations which have been used in this study~\citep{Borhanian:2020ypi}. All angle values are in radians. Some of these sites have not been finalized yet and have been planned/proposed for future detectors.}
\label{table:det-loc}
\end{table*}

While the investigations presented in Sec.~\ref{sec:fiducial} are in the context of a single 3G configuration (CE), a three-detector network of 3G detectors is used in the analyses presented in Sec.~\ref{sec:pop}. Our primary 3G network consists of a detector with CE configuration in the US (LIGO-Livingston site), a detector with ET design in Europe (at the Virgo site), and another CE detector in Australia. We refer to this network as the LAE network of 3G detectors. Additionally, three different 3-detector networks have been used to study the detectability of higher order modes as detectors evolve through LIGO A+, LIGO Voyager, and CE configurations. The sensitivity curve~\citep{Gitlab:Voyager} and locations of these detectors are shown in Fig. \ref{fig:PSDs} and Table \ref{table:det-loc}, respectively. For each detector, we put the lower frequency bound ($f_{\text{low}}$) as 5\,Hz following Chamberlain {\it et al.}~\citep{Chamberlain:2017fjl}.\footnote{Note that in Ref.~\cite{Chamberlain:2017fjl}, authors have used a low-frequency cutoff of 1 Hz for ET configuration; however, we work with a universal low-frequency cutoff of 5 Hz for all detector configurations (LIGO A+, LIGO Voyager, CE, or ET) in this work.}


\section{Detectability of higher modes in 3G detectors}
\label{sec:fiducial}

\subsection{Detection criteria}
\label{sec:det-criteria}

\noindent A robust method to quantify confident detection of a weak gravitational wave signal in noisy detector data involves computing the signal-to-noise ratio (SNR). Assuming, a Gaussian noise, and that template (signal model) is exactly the same as the signal in the detector data, one can define the ``optimal'' SNR ($\rho$) ~\citep{Th300, Cutler:1992tc, Cutler:1994ys} as

\begin{equation}
\rho^2 = (h|h),
\label{eq:opt-snr}
\end{equation}

\noindent where $(.|.)$ denotes the noise-weighted inner product and $h$ represents the GW strain given by Eq.~\eqref{eq:sph}. 

For any two functions $a$ and $b$, the inner product is defined as

\begin{equation}
(a|b) = 2\bigintssss_{0}^{\infty}\frac{\tilde{a}^*(f)\,\tilde{b}(f) + \tilde{a}(f) \tilde{b}^*(f)}{S_h(f)}df\,,
\label{eq:inner-product}
\end{equation}

\noindent where $e{a}(f)$ represents Fourier transform of the function $a$. In the above, $S_h(f)$ denotes the power spectral density of the detector noise and is a measure of noise in the detector (see Fig. \ref{fig:PSDs} for its shape in different detector configurations). 
Following the definition of optimal SNR [given by Eq.~\eqref{eq:opt-snr}], we can quantify the power in higher order modes by defining the optimal SNR tied to individual modes. We define 

\begin{equation}
\rho^2_{\ell m} = (h_{\ell m}|h_{\ell m})=4\bigintssss_{0}^{\infty}\frac{|\tilde{h}_{\ell m}(f)|^2}{S_h(f)}df\,,
\label{eq:opt-snr-hm}
\end{equation}

\noindent where $\tilde{h}_{\ell m}(f)$, analogous to the GW strain in frequency domain, represents strain for any $(\ell, \pm m)$ mode pair and can be expressed as a linear combination of associated polarizations, $\tilde{h}_+^{\ell m}(f)$ and $\tilde{h}_\times^{\ell m}(f)$, as 
 
\begin{equation}
\tilde{h}_{\ell m}(f) = F_+(\theta, \phi, \psi)\,\tilde{h}^{\ell m}_+(f)+F_\times(\theta, \phi, \psi) \,\tilde{h}^{\ell m}_\times(f)
\label{eq:hlmf}
\end{equation}

\noindent where the antenna pattern functions $F_+(\theta, \phi, \psi)$ and $F_{\times}(\theta, \phi, \psi)$ are functions of two angles ($\theta$, $\phi$) giving binary's location in sky and the polarization angle ($\psi$). The two polarizations associated with each $(\ell, \pm m)$ mode pair ($\tilde{h}_+^{\ell m}(f)$, $\tilde{h}_\times^{\ell m}(f)$) can suitably be expressed using a basis of spin-weighted spherical harmonics of weight $-2$ in frequency domain as (see Appendix C of Ref.~\citep{Mehta:2017jpq} for details and the derivation)

\begin{equation}
\label{eq:mode-pol}
\begin{split}
\tilde{h}_+^{\ell m}(f) & = \bigg[(-)^\ell \frac{d_2^{\ell, -m}(\iota)}{d_2^{\ell m}(\iota)} + 1\bigg] Y_{-2}^{\ell m}(\iota, \varphi_0) \tilde{h}^{\rm R}_{\ell m}(f)
\\
\tilde{h}_\times^{\ell m}(f) & = -{\rm i} \bigg[(-)^\ell \frac{d_2^{\ell, -m}(\iota)}{d_2^{\ell m}(\iota)} - 1\bigg] Y_{-2}^{\ell m}(\iota, \varphi_0) \tilde{h}^{\rm R}_{\ell m}(f)
\end{split}
\end{equation}

\noindent where $\tilde{h}^{\rm R}_{\ell m}(f)$ represents the Fourier transform of the real part of the $h_{\ell m}(t)$ appearing in Eq.~\eqref{eq:sph}, $d^{\ell, m}_2(\iota)$ are the Wigner $d$ functions, and $Y^{\ell, m}_{-2}(\iota, \varphi_0)$ are spin-weighted spherical harmonics of weight $-2$ (see for example, Ref.~\cite{Wiaux:2005fm}).\footnote{In writing Eq.\,\eqref{eq:mode-pol} we have set for the spherical angles appearing in Eq.\,\eqref{eq:sph}, $(\Theta, \Phi)\equiv(\iota, \phi_0)$ , where $\iota$ is binary's inclination angle and $\varphi_0$ is a reference phase.} Note that $\tilde{h}^{\rm R}_{\ell m}(f)$ can be expressed in terms of an amplitude and a phase associated with each mode as   

\begin{equation}
\label{eq:hlm-R}
    \tilde{h}^{\rm R}_{\ell m}(f) = A_{\ell m}(f)\,e^{i\varphi_{\ell m}(f)}\,,
\end{equation}

\noindent where $A_{\ell m}(f)$ and $\varphi_{\ell m}(f)$ are obtained in the frequency domain by performing fits to a set of target waveforms chosen appropriately (see for instance Ref.~\cite{Mehta:2017jpq}). Details and the explicit expressions for the amplitude and the phase models used in this work can be found in Eqs.\,(4)-(9) of~\citep{London:2017bcn}.

This definition of mode SNR [given by Eq.\,(\ref{eq:opt-snr-hm})] closely follows the one in Ref.~\cite{Mills:2020thr} which was used for quantifying the SNR of the 33 mode for GW190814~\citep{Abbott:2020khf} and GW190412~\citep{LIGOScientific:2020stg}.  Other methods used for detecting the presence of higher modes have been discussed in \citep{Ghonge:2020suv, Roy:2019phx, OBrien:2019hcj}. For our purposes, we choose to work with a threshold of $3$ on SNR for individual higher modes ($\rho_{\ell m}$) defined above, and of 10 for the dominant 22 mode. This would mean that a confident detection of a source in the dominant 22 mode requires the corresponding SNR to be above 10, and that of a higher mode requires the corresponding SNR to be above 3. 
The choice of the higher mode SNR threshold of $3$ is motivated by the measures adopted in \citep{LIGOScientific:2020stg} that discusses the detection of the 33 mode in the data for the event GW190412.

It is important to note that the observed gravitational waveform is a superposition of different spherical harmonic modes, and hence the total SNR would contain contributions from the {\it interference} terms between different harmonics~\cite{VanDenBroeck:2006ar,Arun:2007qv}. They are likely to contribute negligibly to the total SNR, compared to the dominant contributions (given by  Eq.~\eqref{eq:opt-snr-hm}) as shown in \cite{Mills:2020thr} in the context of aLIGO detectors. Regardless of the magnitude of the interference terms, Eq.~(\ref{eq:opt-snr-hm}) should be seen as a definition of SNR in different modes. 

For all practical purposes we can choose to work with a low- and high-frequency cutoff ($f_{\rm low}, f_{\rm cut}$) and reexpress the optimal SNR for each mode as
\begin{equation}
\rho^2_{\ell m} =4\bigintssss_{f_{\rm low}}^{f_{\rm cut}}\frac{|\tilde{h}_{\ell m}(f)|^2}{S_h(f)}df\,.
\label{eq:opt-snr-hm-fcut}
\end{equation}
As discussed above, we choose a universal lower frequency cutoff of 5Hz following Ref.~\citep{Chamberlain:2017fjl}. The high-frequency cutoff ($f_{\rm cut}$) 
though is decided by the mass of the binary and chosen automatically by waveform module with high enough value so as to not lose any signal power \cite{London:2017bcn}.  

\subsection{Higher modes in the $q-\iota$ plane}
\label{sec:q-iota}

\begin{figure*}[htbp!]%
\includegraphics[trim=40 10 40 10, clip, width=.325\linewidth]{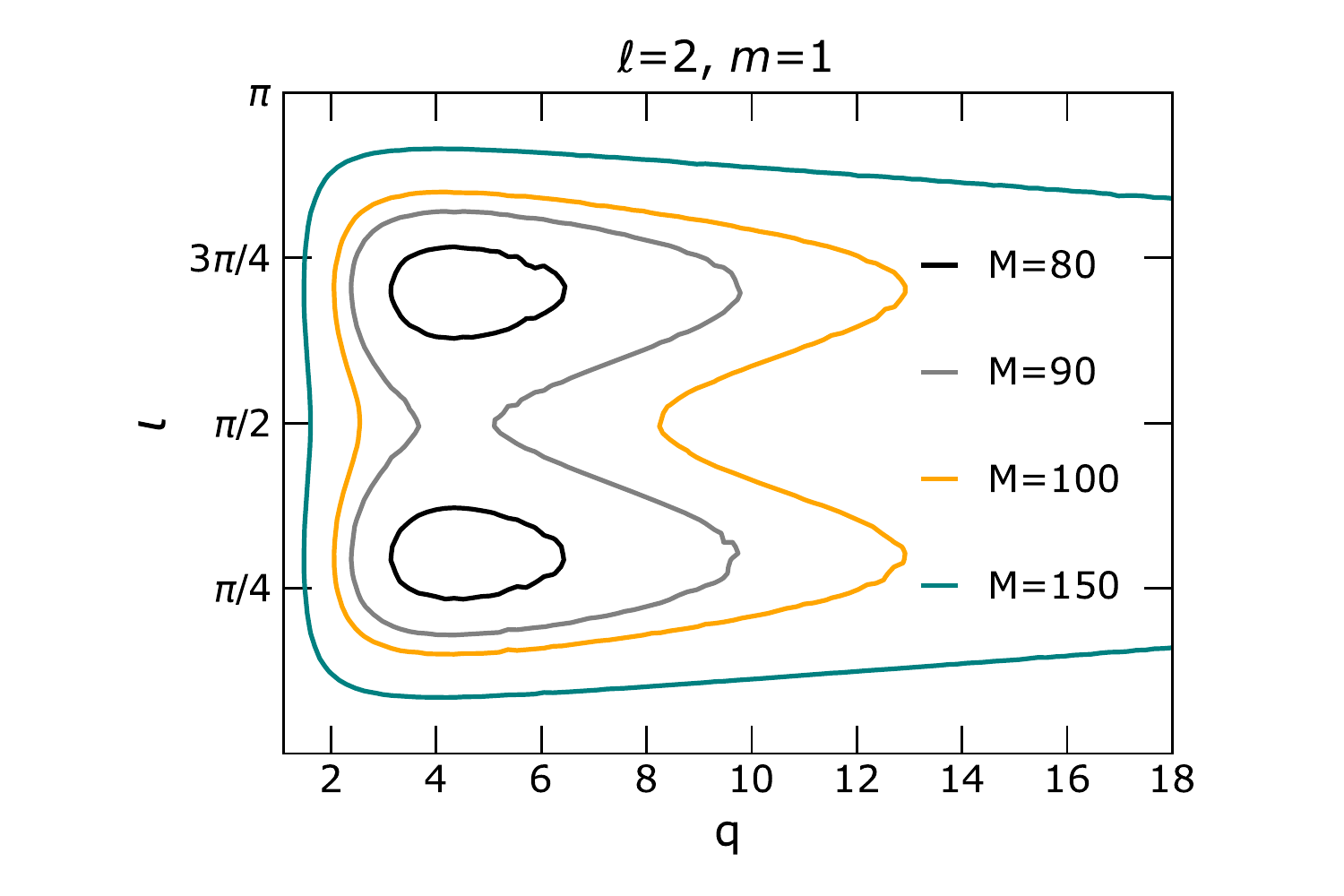}
\includegraphics[trim=40 10 40 10, clip, width=.325\linewidth]{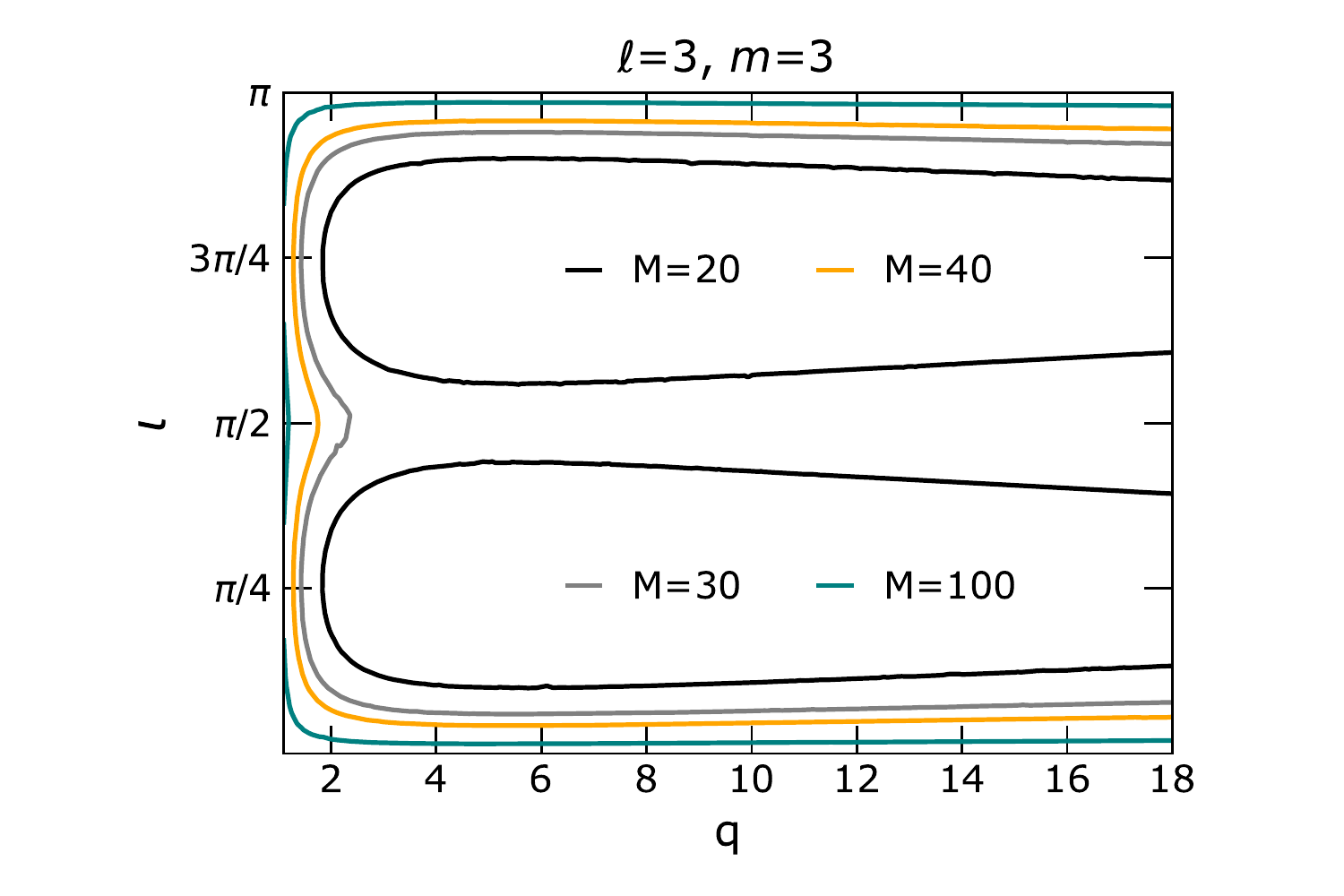}
\includegraphics[trim=40 10 40 10, clip, width=.325\linewidth]{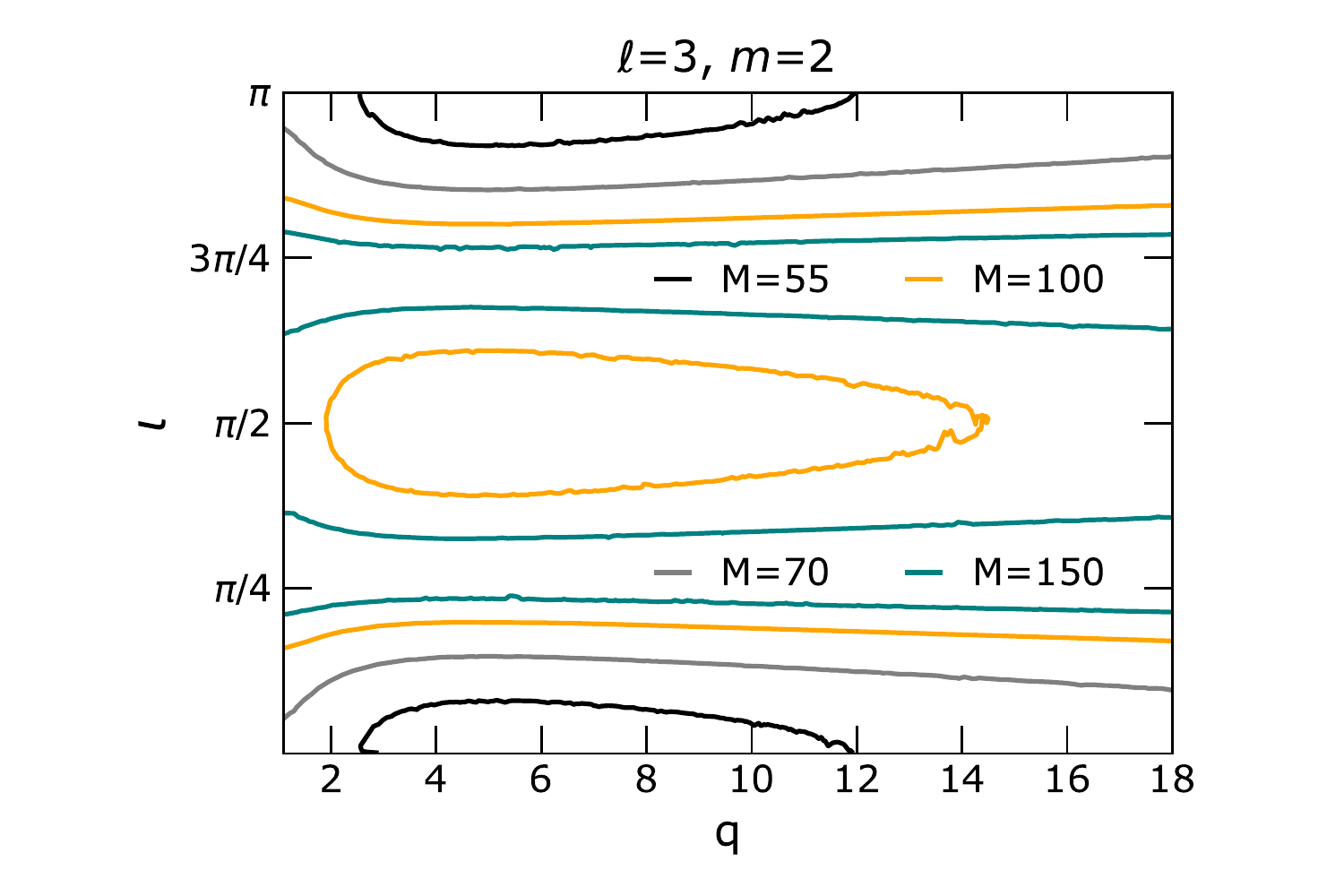}
\includegraphics[trim=40 10 40 10, clip, width=.325\linewidth]{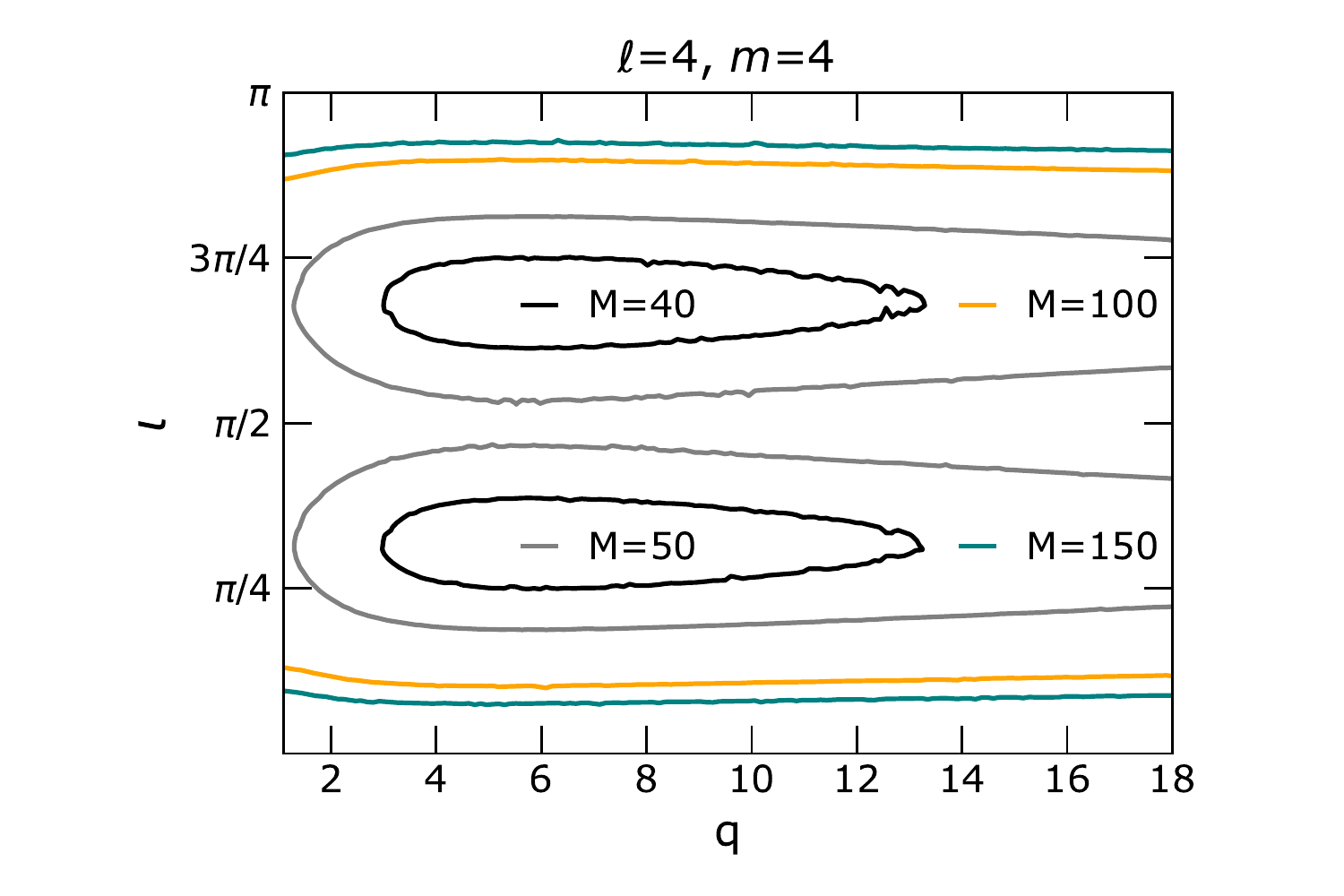}
\includegraphics[trim=40 10 40 10, clip, width=.325\linewidth]{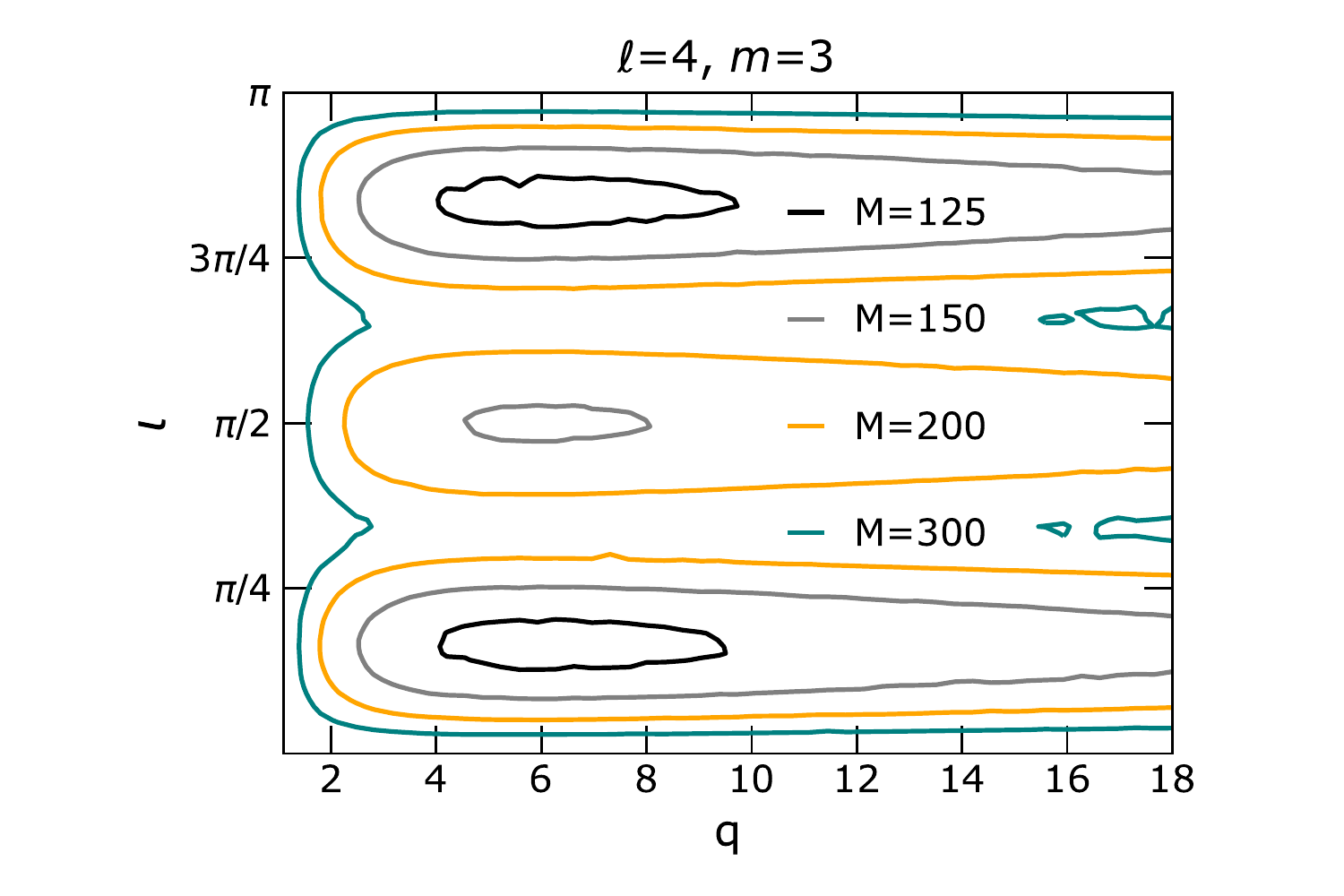}
\caption{Figure shows fixed SNR (=3) contours for various non-quadrupole modes, corresponding to different total mass systems in the $q-\iota$ plane. Each plot corresponds to a particular mode (see panel title). In a particular plot, the contours correspond to different values of total mass. All the systems have been taken at a fixed distance of 3\,Gpc, and with spin values $\chi_{1z}$=0.9 and $\chi_{2z}$=0.8.}
\label{fig:q-iota}
\end{figure*}

Now that we have a formal definition for the optimal SNR for any mode, and have established a detection criterion, in the sections that follow we shall present the result of our investigations. One of the first things that we intend to discuss here concerns the detection of various subdominant modes in the $q-\iota$ plane. Recall, the discussion in Sec.~\ref{sec:intro} that these modes not only become relevant when binary has mass asymmetry ($q>1$) and is not optimally inclined, ($\iota\neq0$) but might also be detected frequently by the next-generation detectors such as those in the 3G era. Hence, quantifying the detectability of higher modes in the $q-\iota$ plane is very important with regard to the physics that is associated with the compact binary mergers. This section focuses on exploring the parameter space in the $q-\iota$ plane which will be accessible through higher order modes in the 3G era. The results of this investigation are summarized in Fig.~\ref{fig:q-iota}.

The analysis here considers the SNR corresponding to a single detector: CE placed at the location L (refer Table \ref{table:det-loc}). Binaries, which act as representative systems, have a fixed value of dimensionless spin components as $\chi_{\rm 1z}$=0.9 and $\chi_{\rm 2z}$=0.8 and are assumed to be kept at a distance of 3\,Gpc, with sky location and polarization angles as $\theta=30^0$, $\phi=45^0$, and $\psi=60^0$, a choice that, although arbitrary, has no impact on the conclusions.

Figure \ref{fig:q-iota} shows the fixed SNR contours for various higher order modes in the $q-\iota$ plane. Each contour corresponds to the fixed (single-detector) SNR of 3 and the region inside each contour has an SNR higher than 3. Contours corresponding to different choices of total mass are displayed in the plot. The contours provide the regions in the $q-\iota$ plane where detection of different subdominant modes will be plausible. In other words, any source which lies inside the contours will be detectable, whereas those which lie outside will not be detectable. We observe that while the detection of 33 mode is possible for masses as low as $20M_{\odot}$, 44 mode can only be detected in heavier systems as displayed by contours in the bottom left panel of Fig.~\ref{fig:q-iota}. This is not surprising since the 44 mode (compared to the 33 mode) is more sensitive to high frequencies. As heavier systems merge at lower frequencies, they bring the higher mode content to the sweet spot of the detector band allowing the accumulation of SNR. Note also, for a given binary the 44 mode amplitudes are relatively lower than the 33 amplitude and more or less increase linearly with its total mass. This naturally affects the power in a given mode and can explain the nondetection of 44 mode in lighter systems. Similar arguments (based on the frequency sweep of each mode in the detector's band and their relative amplitude) can be outlined to explain the trends seen in Fig.~\ref{fig:q-iota} with respect to the minimum mass for which a certain mode is detected.

The trends in mass ratio and inclination angle are distinct for each mode. For a particular total mass value, as $q$ increases, we see that the contours become narrower until they close at a point. Any binary with a mass ratio value higher than this point will not be detectable. The maximum mass ratio for detectable binary is very different for each mode. For a total mass of 100 M$_\odot$, the mass ratio reach of 33, 44, and 32 modes is well beyond 18, whereas for 21 mode it is only up to 13, and for 43 mode, the binary is not even detectable. This is also because the SNR keeps reducing and so the detectability of HMs also reduces. But, as we have discussed earlier, the relative contribution of HMs increases as we go to higher values of $q$. 

We can see these distinct (and somewhat complementary) trends in $\iota$ as well. Again, for a total mass of 100 M$_\odot$, and a fixed mass ratio (say $q$=10), 33 mode covers almost the entire $\iota$ range whereas for 44 mode it is somewhat restricted. It is interesting to note that for 21 and 32 modes, the $\iota$ coverage is almost complementary; with 21 covering (a little more than) the range between ($\pi/4$, $3\pi/4$) and 32 covering the rest. It can also be seen that as the total mass is increased, the contours include a larger region of the parameter space. The bi- and trimodality of these contours reflect the symmetries these modes possess with respect to change in $\iota$. Though we have only shown the results of the high-spin case here, the trends remain the same for low spins too. Going from low spins ($\chi_{1z}=0.3$, $\chi_{2z}=0.2$) to high spins increases the SNR very slightly in 33 mode (less than 15\% increase), moderately in 44 and 32 modes (nearly 30\% increase), and visibly in 21 and 43 modes (nearly 50\% increase). This however, does not change the overall shape of the contour.

\subsection{Detecting higher modes of GWTC-2 events using 3G detectors}
\label{sec:gwtc2}

Next, we investigate the detectability of higher modes for a few selected GWTC-2 catalog events assuming the sensitivity of 3G detectors. This helps us assess the improved detection rates that can be expected due to the use of an advanced detector configuration over what we already have from the present detectors. For this, we have chosen a few representative events from the GWTC-2 catalog \citep{LSC:GWTC-2-GWOSC} with high detection significance, either because higher modes have already been detected for them by the LIGO-Virgo observations (GW190814, GW190412), or because the inclusion of higher modes in the waveform significantly improved the parameter estimation of the events~\citep{LIGOScientific:2020ibl}. 

Again, in this analysis, we consider the SNRs corresponding to a single CE detector placed at the location L (see Table \ref{table:det-loc}). For each event, in order to compute the distribution of SNRs for different higher order modes, we take 10,000 random posterior samples from the corresponding dataset available for that event. (see Ref.~\citep{LSC:GWTC-2-GWOSC} for complete datasets). We then take the median value of SNR from this distribution of 10,000 points and quote this value for each mode in Table \ref{table: GWTC-2 events}.

As expected, there is a significant improvement in the detection rate of higher modes as compared to the aLIGO and Virgo with their current sensitivities. The single 3G detector shows promise of detecting 33, 44, 21, and 32 modes for all of the above-mentioned events. It is noteworthy that GW190814 shows the highest SNR values for the higher modes, as well as the highest relative SNR for 33 mode. This can clearly be explained by the high mass ratio value ($q \sim 9$) of this event. The 33 mode network SNR reported by LIGO-Virgo for this event was $\sim$6.6, whereas we can see that for a single CE detector, this number becomes $\sim$170. Similarly, the relative contribution of higher modes is considerable (by a factor larger than 20) for GW190412. Such high SNR can be attributed to the event's mass ratio ($q \sim 3.2$) and nonzero effective spin.

To summarize, several of the GWTC-2 events would have led to reliable detection of all the four leading modes of gravitational waveforms with the sensitivities of the proposed 3G detectors. However, a robust method to quantify the detectability of higher modes must involve synthesizing a population based on the inferences from the LIGO/Virgo detections so far. This forms the theme for the next section.

\begin{table}[]
\begin{tabular}{|c|c|c|c|c|c|c|c|c|}
\hline
\multirow{2}{*}{\textbf{Event}} &
  \multirow{2}{*}{\textbf{M}} &
  \multirow{2}{*}{\textbf{q}} &
  \multirow{2}{*}{\textbf{$\chi_{\text{eff}}$}} &
  \multicolumn{5}{c|}{\textbf{SNR in mode}} \\ \cline{5-9} 
                 &        &      &      & \textbf{22} & \textbf{33} & \textbf{44} & \textbf{21} & \textbf{32} \\ \hline
GW190412         & 42.6  & 3.2 & 0.2 & 649      & 81        & 17       & 14        &    3.9    \\ \hline
GW190519\_153544 & 159.5 & 1.6 & 0.4 & 685      & 79       & 48       & 19       & 14       \\ \hline
GW190521         & 279.8 & 1.4 & 0.1 & 424      & 22       & 19       & 7.1        & 7.5        \\ \hline
GW190602\_175927 & 173.8 & 1.4  & 0.1 & 330      & 15       & 9.4        &  4.3      &    4.7   \\ \hline
GW190630\_185205 & 69.9   & 1.5 & 0.1  & 708      & 31       & 14       & 7.9        & 5.8        \\ \hline
GW190706\_222641 & 183.5 & 1.7 & 0.3 & 223      & 18       & 7.6        &  3.4      & 3.3       \\ \hline
GW190814         & 27.2  & 9.0 & 0    & 982      & 172      & 33       & 32       &   4.4      \\ \hline
GW190828\_065509 & 44.4  & 2.4 & 0.1 & 418      & 34       & 7.2        & 6.7        &  3.0      \\ \hline
\end{tabular}
\caption{Detectability of higher modes in GWTC-2 events using a 3G detector. We have sampled the parameter values from the posteriors of these events\citep{Abbott:2020tfl, Abbott:2020khf, LIGOScientific:2020stg, LIGOScientific:2020ibl, LSC:GWTC-2-GWOSC}, and have quoted the median values of SNRs obtained from 10,000 posterior samples. The total mass values quoted above are the detector frame masses. While in GWTC-2, only GW190412 and GW190814 showed a detection of the 33 mode, it can be seen in the above table that many more events would have shown detectability of HMs in a 3G detector configuration.}
\label{table: GWTC-2 events}
\end{table}
\section{Population Study}
\label{sec:pop}

Our knowledge of the BBH population in the universe has evolved from the first observing run through the first half of the third observing run of the LIGO/Virgo detectors. Here, we employ the state-of-the-art population model of Ref.~\citep{LIGOScientific:2020kqk} to synthesize a BBH population and assess the detectability of various subdominant modes by using the method introduced earlier. 

The details of these populations are discussed next.

\subsection{Population models}
\label{sec:pop models}

\begin{figure*}[htbp!]%
\includegraphics[trim=10 10 10 10, clip, width=.45\linewidth]{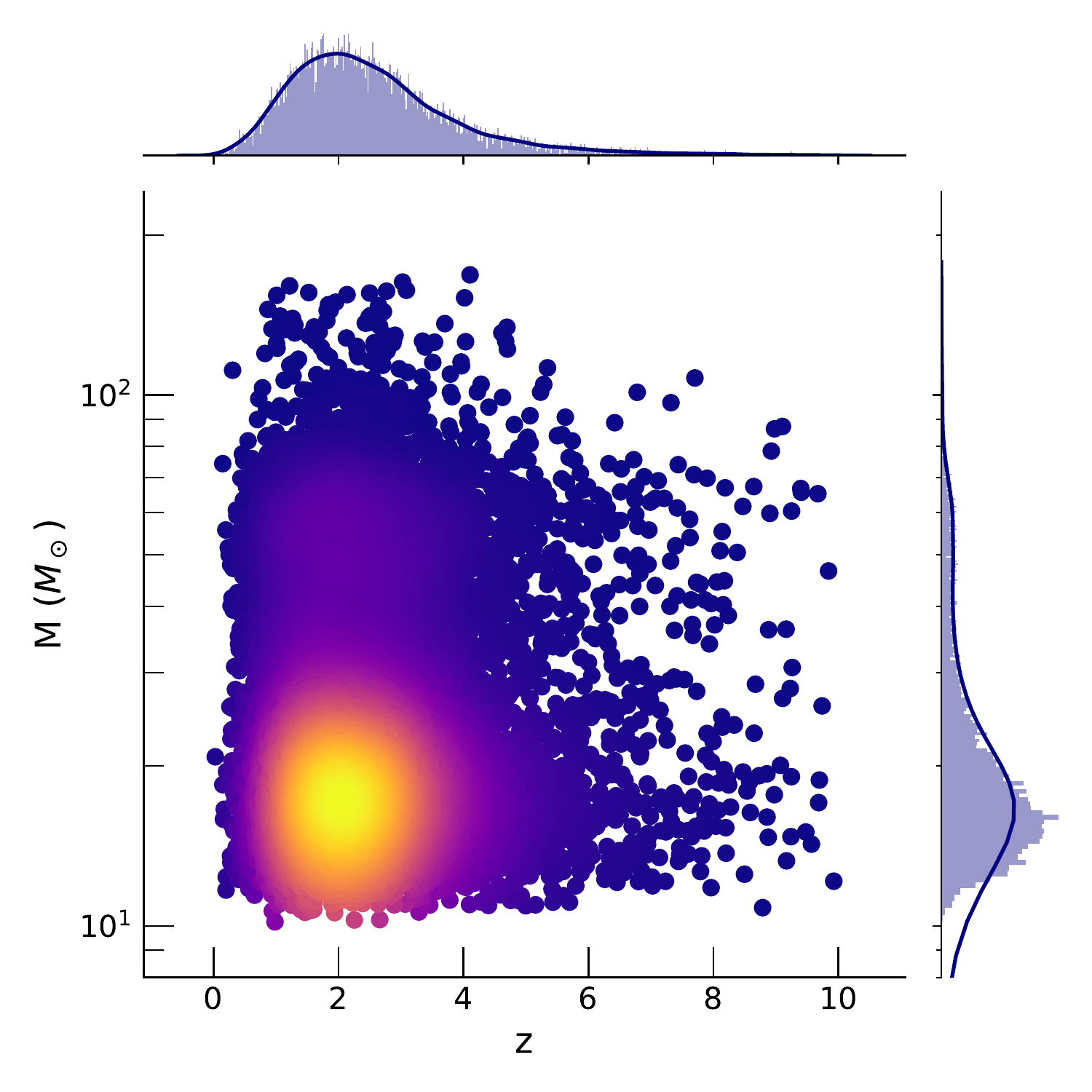}
\includegraphics[trim=10 10 10 10, clip, width=.45\linewidth]{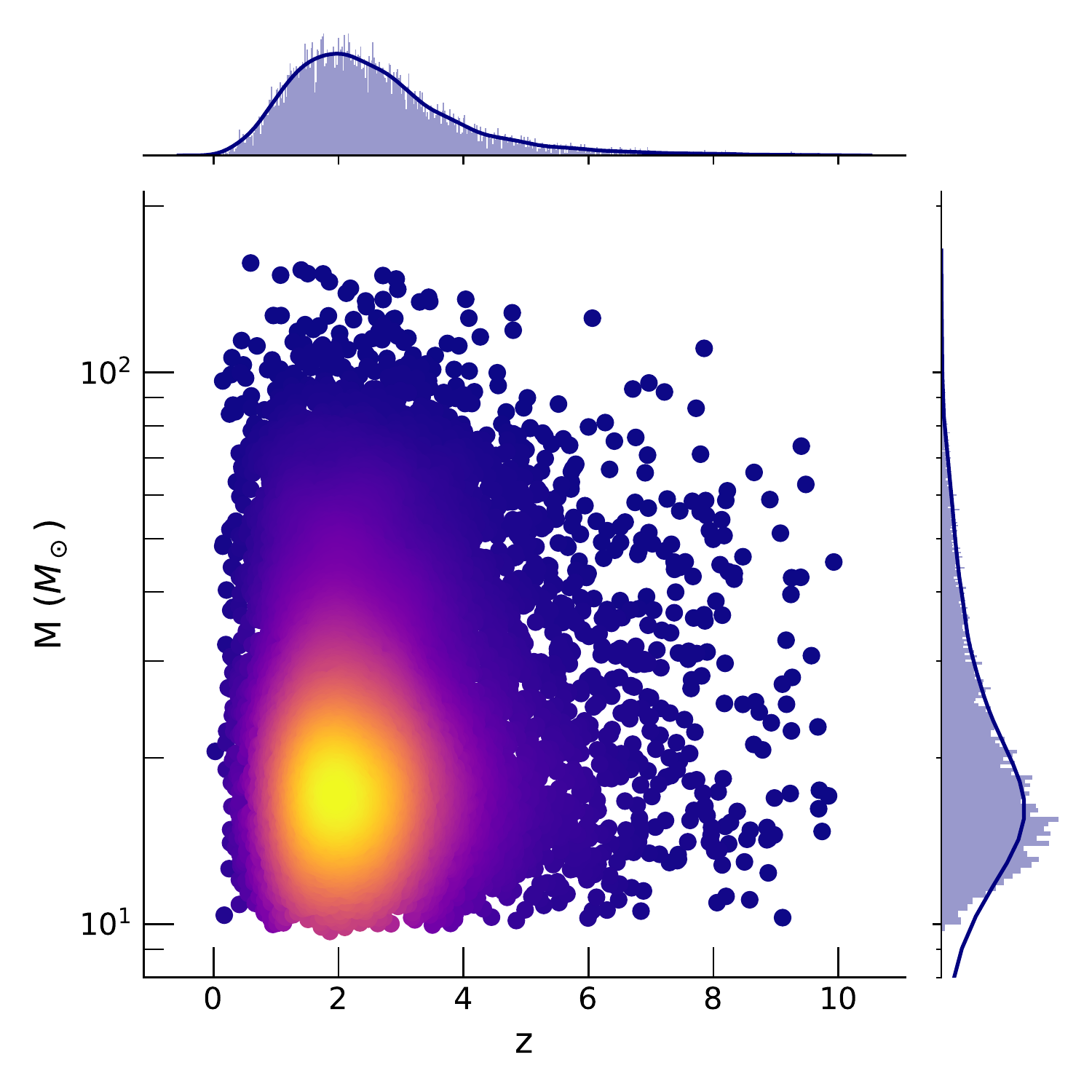}
\caption{Figure shows the 2D distribution of injected values of total mass ($M$) and redshift($z$). Left: Mass model is Power Law + Peak, Right: Mass model is Broken Power Law. Redshift has been distributed according to the MDBN merger rate, and has been taken to be same for both the mass distributions.}
\label{fig:mass-models}
\end{figure*}

\begin{table}[]
\begin{tabular}{|c|c|c|c|}
\hline
\multicolumn{2}{|c|}{\textbf{Power Law + Peak}} & \multicolumn{2}{c|}{\textbf{Broken Power Law}} \\ \hline
Parameter        & Value & Parameter  & Value \\ \hline
$\alpha$         & 2.63  & $\alpha_1$ & 1.58  \\ \hline
$\mu_m$          & 33.07 & $\alpha_2$ & 5.59  \\ \hline
$\sigma_m$       & 5.69  & b          & 0.43  \\ \hline
$\delta_m$       & 4.82  & $\delta_m$ & 4.83  \\ \hline
$\beta$          & 1.26  & $\beta$    & 1.4   \\ \hline
$m_{min}$        & 4.59  & $m_{min}$  & 3.96  \\ \hline
$m_{max}$        & 86.22 & $m_{max}$  & 87.14 \\ \hline
$\lambda_{peak}$ & 0.10  &            &       \\ \hline
\end{tabular}
\caption{Values of model parameters for mass models used in the population study.}
\label{table: pop models}
\end{table}
We consider two different mass distribution models, the Power Law + Peak (PL+P) and Broken Power Law (BPL) outlined in Ref.~\cite{LIGOScientific:2020kqk}. The primary mass distribution for the PL+P model is given by

\begin{equation}
p(m_1) = \big[(1-\lambda_{\rm peak})\mathcal{B}(m_1) + \lambda_{\rm peak}G(m_1)\big]S(m_1)
\end{equation}
where
\begin{equation}
\begin{aligned}
\mathcal{B}(m) &= \mathcal{C}m^{-\alpha},\ \ m < m_{max},\text{\ \ \ \ and}\\
G(m) &= \frac{1}{\sqrt{2\pi}\sigma_m}\big[e^{-(m-\mu_m)^2/2\sigma_m^2}\big]
\end{aligned}
\end{equation}

Here $\mathcal{C}$ is a normalization constant, and $S(m_1)$ is the smoothing function given by Eq.~(B6) of \cite{LIGOScientific:2020kqk}. The mass ratio for both mass distribution models (PL+P and BPL) is given by a power law that also includes the smoothing term and is given as
\begin{equation}
    p(q) = q^\beta S(m_1 q)
\end{equation}
For the BPL model the primary mass ($m_1$) is distributed as follows

\begin{equation}
p(m_1) \propto 
\begin{cases}
m_1^{-\alpha_1}S(m_1), & m_1 < m_{\rm break}\\
\\
m_1^{-\alpha_2}S(m_1), & m_1 > m_{\rm break}\\
\\
0, & \text{otherwise}
\end{cases}\\
\end{equation}
where $m_{\rm break} = m_{\rm min} + b(m_{\rm max} - m_{\rm min})$.

The values of hyperparameters used in the above-mentioned models are given in Table \ref{table: pop models}. We have distributed the primary mass ($m_1$) in the limit of [5, 100]$M_\odot$ and the mass ratio ($q=m_1/m_2$) in the range of [1, 18], which is the maximum $q$ up to which the waveform used here is calibrated \cite{London:2017bcn}.

To distribute these sources to redshifts accessible to a 3G detector network, we have employed the Madau-Dickinson-Belczynski-Ng model for field BBH merger rate of Ref.~\cite{Ng:2020qpk}. The volumetric merger rate reads

\begin{equation}
    \dot{n}_{\rm F}(z) \propto \frac{(1+z)^{ \alpha_{\rm F}}}{1+\big[ (1+z)/C_{\rm F} \big]^{\beta_{\rm F}}}
\end{equation}
with $\alpha_{\rm F} = 2.57, \beta_{\rm F} = 5.83, C_{\rm F} = 3.36$. Further details regarding this can be found in Appendix~B of \cite{Ng:2020qpk}. Using this model for the merger rates, the redshift is then distributed as follows:\footnote{We have used the recently developed \textsc{python} package \textsc{gwbench}~\cite{Borhanian:2020ypi} for the distribution of redshift.}

\begin{equation}
    p(z) \propto \frac{4\pi\dot{n}_F(z)}{1+z} \left(\frac{dV_{\rm c}}{dz}\right)
\end{equation}
where $V_{\rm c}$ represents the comoving volume. The range for $z$ has been taken as [0, 10]. The redshift distribution of population sources, along with the resultant (total) mass distribution from the PL+P and BPL mass models is shown in Figure \ref{fig:mass-models}. 

The spins are distributed using the Default Model of \cite{LIGOScientific:2020kqk} for dimensionless spin magnitude ($\chi_{1,2}$) given by

\begin{equation}
p(\chi) = \text{Beta}(\alpha_\chi, \beta_\chi).
\label{eq: spin mag}
\end{equation}

\noindent 
The values of $\alpha_\chi$ and $\beta_\chi$ are computed using the mean ($\mu_\chi$) and variance ($\sigma^2_\chi$) for the Beta distribution. We found these values to be: $\alpha_\chi=$ 6.3788 and $\beta_\chi=$ 2.2412, and used them in constructing the distribution for dimensionless spin magnitudes ($\chi_{1,2}$). Further the cosine of the tilt angle, defined as $z_\chi = \cos(\theta_{1,2})$, is distributed as $p(z_\chi)$ (see Sec. D1 of Ref.~\citep{LIGOScientific:2020kqk} for related details). The distribution $\chi_{\rm 1z}$ and $\chi_{\rm 2z}$ reads
\begin{equation}
p(\chi_{\rm 1z,2z}) = p(\chi_{\rm 1z,2z})\ p(z_\chi).
\label{eq:spin-dist}
\end{equation}
\noindent We have taken the same distribution for $\chi_{\rm 1z}$ and $\chi_{\rm 2z}$, in the range [-1,1].\footnote{Note that the waveform we use (\texttt{IMRPhenomHM}) is calibrated up to spin magnitude of 0.85 (0.98 for equal mass systems); however, we have extended this up to 1 in order to include the complete range of spin magnitudes.}

\subsection{SNR distribution of higher modes}
\label{sec:pop hist}

We simulate 10,000 sources following the above-mentioned prescription. We vary all the nine parameters, the luminosity distance ($D_L$), inclination angle ($\iota$), total mass ($M$), mass ratio ($q$), sky angles ($\theta, \phi$), orientation angle ($\psi$), and spins ($\chi_{1z}$ and $\chi_{2z}$), following the population models. The ranges for the total mass, mass ratio, spins, and redshift have been mentioned with their respective population models in Sec. \ref{sec:pop models}. The cosines of the angles $\iota$ and $\theta$ have been varied uniformly between (-1, 1), and $\phi$ and $\psi$ are uniform between (0, 2$\pi$). It is worth mentioning that due to the low mass of the secondary, which falls in the NS-BH mass gap, GW190814 is an outlier. Thus, the O3a population models do not include it while calculating the hyperparameters \citep{LIGOScientific:2020kqk}.

\begin{figure}%
\centering
\includegraphics[trim=30 0 30 30, clip, width=\linewidth]{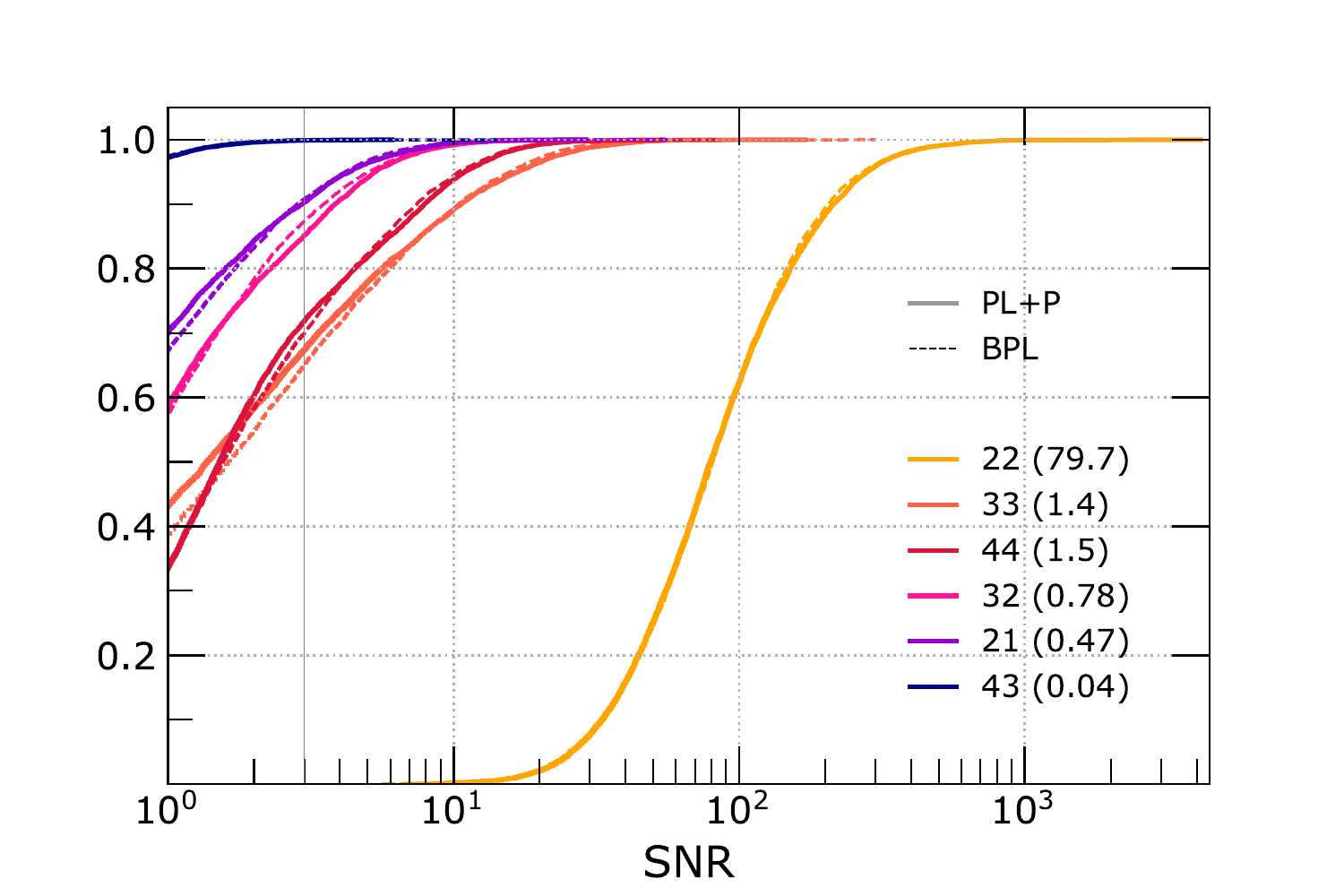}
\caption{Cumulative SNR histograms for six leading modes, with the two mass distribution models (PL+P and BPL) coupled with the MDBN model for the redshift evolution shown. The plot shows the trends for each mode for SNRs $>1$. The solid and dashed colored lines correspond to population drawn from the PL+P and BPL models, respectively. The values written in the brackets are median values of SNR for each mode. The detector network used here consists of a detector with CE configuration in the U.S. (at LIGO-Livingston site), a detector with ET design in Europe (at the Virgo site), and another CE detector in Australia and referred to as 3G LAE network in this work.} 

\label{fig:SNR histograms}
\end{figure}

Figure \ref{fig:SNR histograms} shows the cumulative histograms for the SNRs of various modes in the LAE network (two CE detectors, one ET detector) of 3G detectors (see Sec.~\ref{sec:det-network} for details). The numbers in brackets are the median values of SNR for each mode. For a particular value of SNR (on the $x$ axis), the $y$ axis corresponds to the fraction of population having SNR up to that value. We find that 99.8\% of the simulated population following the PL+P model have SNR $>$10 in 22 mode. We show the results for both the populations. The dashed lines correspond to the population distributed according to the BPL mass model, while the solid lines correspond to the PL+P distribution. Since the results from the two models are very close,  we only quote numbers corresponding to the PL+P model. Note that the detection fractions that we quote in the subsequent sections are obtained by putting a minimum cutoff of SNR $>10$ for the 22 mode, and SNR $>3$ for all the other modes.  

We find that the 33 mode is detectable in nearly 33\% of the sources and 44 mode is detected in $\sim$28\% of the population. 32 and 21 modes are detectable in nearly 15\% and 10\% sources, respectively, while the detected fraction is only $\sim$0.1\% for the 43 mode. This demonstrates that for a population of sources similar to those detected in GWTC-2, with the increased reach of 3G detectors to higher redshifts, we will detect most higher modes in a large number of sources.

\subsubsection{Comparison between various generations of detector networks}
\label{subsec:2-2.5-3G hist}

\begin{figure}%
\centering
\includegraphics[trim=30 0 30 30, clip, width=
\linewidth]{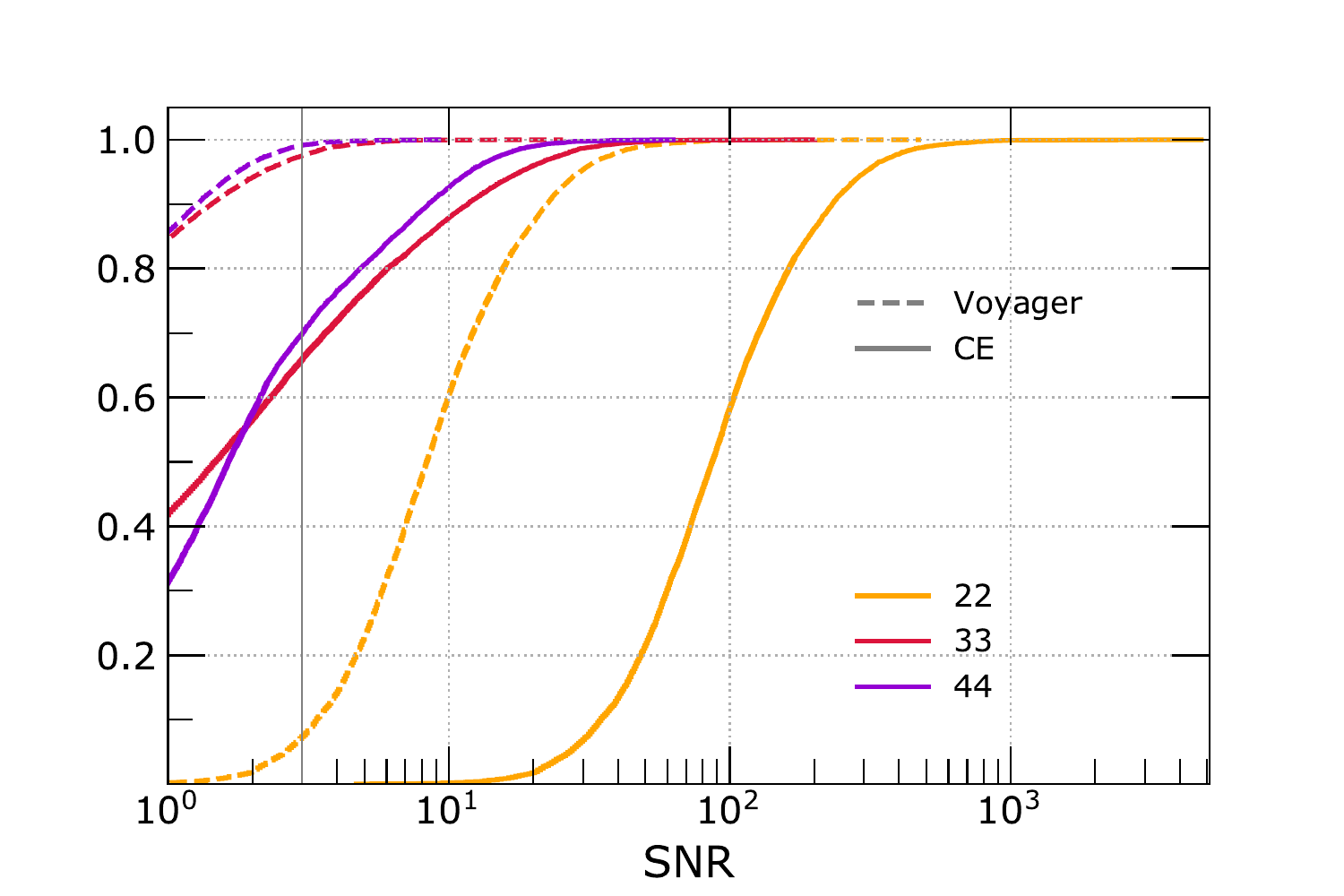}
\caption{Cumulative histograms of SNR for various higher modes as a comparison between various generations of detectors. The solid colored lines denote 3G detector network with CE, while the dashed lines denote the detector network using LIGO Voyager. Both the detectors have been placed at the locations of the LHV network. We have taken $q_{max}=18$. 
}
\label{fig:SNR 2.5-3g}
\end{figure}

In this section, we investigate how the detectability of higher modes changes for a population across various generations of detector networks. We consider three kinds of detector networks for this study: LIGO-A+, LIGO-Voyager, and CE detector networks. In order to avoid detector location bias, we have chosen the same set of locations, Livingston and Hanford in the U.S. and Cascina in Italy (LHV) for all three networks. {For constructing this population, we have considered the Power Law + Peak mass model.} 

We have shown the results for CE and Voyager networks in Fig.~\ref{fig:SNR 2.5-3g}. Even a quick look at Fig.~\ref{fig:SNR 2.5-3g} reveals that the LIGO Voyager detector network can barely detect the two additional modes besides the quadrupole mode, whereas 3G detectors have a good detection percentage for 33 and 44 modes. As expected, the improvement with 3G detectors is highly significant compared to the upgraded 2G configurations. Note that here we quote the detection percentages of higher modes, over and above the detection of 22 mode. Therefore, the numbers quoted here can be interpreted as the percentage of sources which will show HMs provided the signal has already been detected. 

Again, as seen in the previous section, nearly 100\% sources are detectable in 22 mode in the detector network formed by the CE detector at locations of current LIGO and Virgo detector cites. Out of these, nearly 35\% and 30\% of the sources show a detection in 33 and 44 modes, respectively. This detection percentage drastically decreases for Voyager and A+. For Voyager, only 40\% of the simulated population is detectable in 22 mode, out of which only $\sim$6\% and $\sim$2\% show a detection in 33 and 44 modes, respectively. For A+, only about 7\% of the systems show a detection in 22 mode, out of which $\sim$1\% show 33 mode, and $\sim$0.5\% show a detection in 44 mode. Considering the detection rate numbers from \cite{Baibhav:2019gxm}, these percentages can still result in the detection of a considerable number of HMs in the detected population. 

This shows the tremendous potential of a network of a 3G detectors, and how the number of higher mode detections will significantly rise compared to the currently operating 2G detectors. This, in turn, will also have a profound impact on the overall parameter estimation capabilities of the 3G detectors, and hence influence the astrophysics and fundamental physics in the 3G era in a big way.

\section{conclusions} 
\label{sec:concl}

We have investigated the detectability of nonquadrupole modes of gravitational wave radiation in mass ratio and inclination angle space (Sec. \ref{sec:q-iota}), and for a few events from the GWTC-2 catalog assuming as if they were detected during the 3G era (Sec. \ref{sec:gwtc2}). We find that various modes activate different regions of the $q$-$\iota$ plane, and show various symmetries in $\iota$ leading to bi- and trimodality in the contours. For the GWTC-2 events, we observe a massive improvement of SNR (by a factor larger than 20 times) for the events which had been reported to show detection in 33 mode by LIGO/Virgo observations (GW190814, GW190412). In the 3G era, events like these promise to show a detection in other higher harmonics as well. Apart from this, we also see detectable SNRs in higher modes for other GWTC-2 events, which emphasizes that the number of events that permit the measurement of higher modes will also increase in the 3G era.

We also performed a population study for 10,000 sources which will be detected by the 3G network, and quote the fraction of population which will show the presence of higher modes (Sec. \ref{sec:pop hist}). It is found that nearly 33\% and 28\% of the sources will have detectable SNRs in 33 and 44 modes, respectively, and other modes will also be detectable in a small percentage of the population. Additionally, we compared this fraction with the fraction of higher modes detectable in the upgraded 2G gravitational wave detector networks such as LIGO A+ and LIGO Voyager, using the PL+P mass model. We conclude that this fraction significantly increases from less than 6\% to nearly 35\% for the 33 mode, with the 3G network.

All the above-mentioned investigations were performed using a spinning inspiral-merger-ringdown higher mode waveform of the Phenom family (\texttt{IMRPhenomHM}), and using a different waveform may alter the numbers only slightly. Effects of the mass ratio and inclination angle were observed in the detectability of higher order modes. The effect of increasing total mass (in discrete values) was also noted in Sec. \ref{sec:q-iota}. While we have shown the detectability of various modes for spinning systems, a more detailed study can be done by including the effect of precession along with the higher modes. Further, as a follow-up, we plan to explore the effect of higher modes in the error analysis of various parameters in the 3G detector era.

\acknowledgments

We thank Ajit Mehta for the review of our manuscript and for offering useful comments. We thank B.S. Sathyaprakash and Anuradha Gupta for useful conversations.\,We are grateful to Ssohrab Borhanian, Sumit Kumar, and Collin Capano for their help with the antenna pattern functions. We thank Sebastian Khan for clarifications on the waveform. We acknowledge the help from users on ``\texttt{pycbc-help} slack channel'' in implementing the PyCBC code. K.G.A. is partially supported by a grant
from the Infosys Foundation. K.G.A. acknowledges the Swarnajayanti Grant No. DST/SJF/PSA-01/2017-18 of DST-India, the Core Research Grant No. EMR/2016/005594, and MATRICS Grant No. MTR/2020/000177 of SERB. The population studies reported in this paper were carried out using the Powehi workstation in the Department of Physics, and AQUA cluster at IIT Madras.

\bibliographystyle{apsrev4-1}
\bibliography{ref-list}

\end{document}